\documentstyle[prl,aps,epsf,multicol]{revtex}

\begin{document}
\draft

\title{Deformation of Quantum Dots in the Coulomb Blockade Regime}
\author{G. Hackenbroich$^{**}$, W.D. Heiss$^*$ and H.A. Weidenm\"uller$
^{*\dagger}$}
\address{$^{**}$Department of Applied Physics, Yale University, 
New Haven, CO, USA, \\
$^*$Department of Physics, University of the
Witwatersrand, Johannesburg,  South Africa, \\
$^{\dagger}$ Max-Planck-Institut f\"ur Kernphysik, Heidelberg,
Germany} 
\date{\today}
\maketitle

\begin{abstract}
We extend the theory of Coulomb blockade oscillations to quantum 
dots which are deformed by the confining potential. We show that shape
deformations can generate sequences of conductance resonances which
carry the {\it same} internal wavefunction. This fact may cause strong
correlations of neighboring conductance peaks. We demonstrate the
relevance of our results for the interpretation of recent experiments
on semiconductor quantum dots.   
\end{abstract}

\begin{multicols}{2}

\narrowtext Recently two experimental groups have reported the
first measurements of the distribution of Coulomb blockade peak
heights in semiconductor quantum dots \cite{Chang,folk}. These
measurements were designed to test various predictions of the
statistical theory of the Coulomb blockade \cite{Jalabert}. This
theory was formulated by Jalabert, Stone and Alhassid for the low
temperature regime $k T < \Delta$, where $\Delta$ is the mean
resonance spacing. In this regime, transport through the quantum dot
occurs by resonant tunneling through a single quasibound state. On the
basis of a random matrix assumption for the quasibound eigenfunctions
of the dot, the statistical distribution of the peak heights of the
conductance resonances was derived \cite{Jalabert} both for systems
with time-reversal symmetry (vanishing magnetic field) and for systems
with broken time-reversal symmetry (with sufficiently large magnetic
field).

In the experiments \cite{Chang,folk}, good agreement was found with
the predicted form of the distribution functions. At the same time,
unexpected correlations were observed \cite{folk} between neighboring
peak heights. The correlations comprised up to four adjacent peaks.
Even stronger correlations have recently been found in a set of
experiments \cite{schu} addressing the phase of the transmission
amplitude through a quantum dot. In these experiments, more than 10
consecutive conductance peaks displayed the same phase, and nearly the
same peak height. This phenomenon has not been explained so far.

In this paper we propose a novel mechanism which can account for the
above-mentioned correlations. It applies whenever the same set of
capacitively coupled gate electrodes is used to define the depth {\em
and shape} of the dot confining potential. This is the case in the
experiments of Refs.\ \cite{Chang,folk,schu} but not for others using
a back gate. We study the effect of a deformation of the potential on
the eigenvalues and eigenfunctions of the dot.

Several authors \cite{byl} have shown that a parabolic potential is a
useful approximation to the confining potential of a quantum dot. To
simplify the argument, we start with this case. We identify sequences
of conductance resonances which, as a result of level crossings, carry
the {\em same} internal wave function. Similar sequences are found if
the parabolic potential is replaced by a more realistic confinement
potential with level repulsion and chaotic classical electron motion.
We first disregard the charging energy of the dot but show later that
our conclusions remain essentially unchanged when this energy is
included.

%\vspace*{-0.8cm}
\begin{figure}
\epsfxsize=2.7in
\centerline{
\epsffile{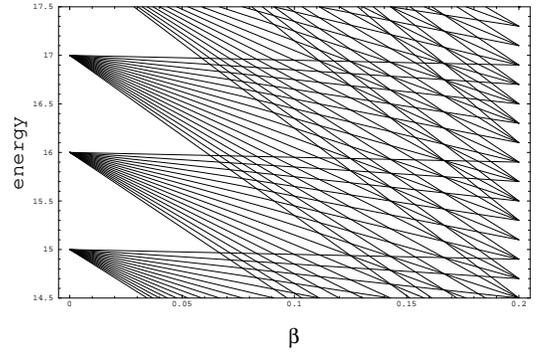}
%\epsffile{wei1.eps}
}
\vglue 0.15cm
\caption{
Energy levels of the two-dimensional harmonic oscillator with
frequencies $\omega_x=1$ and $\omega_y=1-\beta$ as functions of
$\beta $.}
\label{fig1}
\end{figure}

We choose the $x$ axis through the center and in the direction of
transmission through the dot. To account for anisotropies imposed by
the external gate potential (gp), we choose different values for the
frequencies $\omega_x, \omega_y$ of the parabolic confining potential. 
For simplicity, we keep the frequency $\omega_x = \omega^0$ fixed and
change the frequency $\omega_y = \omega^0 (1 - \beta)$, with $0 \leq
\beta < 1$. Increasing $\beta$ widens the potential in the
$y$-direction and thus deforms it. Fig.~1 shows part of the
single-particle spectrum versus $\beta $. The shell structure at
$\beta = 0$ is clearly visible. The energy levels form a network of
intersecting straight lines. Some lines run nearly parallel to the
$\beta$-axis while others have large negative slopes. The explanation
is simple. Each single-particle state is characterized by two quantum
numbers ($n_x,n_y$). States with small (large) values of $n_y$ have
small (large) negative slopes. When two lines cross, the quantum
numbers are carried along the straight lines. We show below that the
pattern in Fig.~1 is generic and applies also to the single-particle
spectrum of deformed non-integrable potentials, with one difference.
The points of intersection of the straight lines disappear and are
replaced by avoided (Landau-Zener) crossings.

To present the essence of our argument, we replace Fig.~1 by an
idealized picture (Fig.~2): A set of equally spaced straight lines
($A$-levels) runs parallel to the $\beta$-axis, a second similar set
($B$-levels) has large negative slope. To model the generic case, we
have replaced actual crossings by avoided crossings. Single-particle
wave functions retain their identity across avoided crossings in
nearly the same way as if the lines were to intersect. Therefore, the
wave function on any $A$-level is nearly independent of $\beta$, and
so is the Fermi energy $E_F$ marked by a dashed line. At a
fixed value of $\beta$, all single-particle states up to $E_F$ are
filled. A change of the gp (and, therefore, of $\beta$) is slow on the
scale of characteristic times of the quantum dot. Therefore, electrons
in occupied orbitals follow the deformation $\beta$ adiabatically. Let 
$\delta \beta$ be the distance between $B$-lines, and gp$_{\delta
\beta}$ be the change of the gp needed to change $\beta$ into $\beta +
\delta \beta$. What happens as we increase the gp by gp$_{\delta
\beta}$? The last occupied $A$-level moves adiabatically one ``floor'' 
down to its nearest analogue, thereby changing its wave function. The
$A$-level right below $E_F$ becomes empty. If gp$_{\delta \beta} =
$gp$_e$, the change of the gp needed to pull another electron into the dot,
then the next conductance resonance will carry the very same
single-particle state as its predecessor. (The Figure simplistically 
suggests that the next resonance occurs whenever a solid line crosses 
$E_F$; in reality the criteria given in \cite{been} apply).
This process can repeat itself for a second,
third, etc.\ time. Thus, subsequent conductance peaks might not be
independent, but would be manifestations of one and the same
single-particle state.  

\vspace*{-0.1cm}
\begin{figure}
\epsfxsize=2.8in
\centerline{
\epsffile{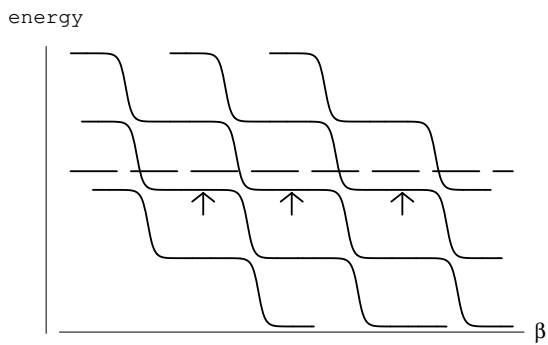}
%\epsffile{wei2.eps}
}
\vglue 0.2cm
\caption{
Schematic illustration of the mechanism described in the text. The
arrows indicate the levels which become available for occupation when
the deformation is increased. 
}
\label{fig2}
\end{figure}

This picture raises several questions, and we devote the remainder of
this paper to the answers. (i) How realistic is Fig.~2? (ii) Does a
significant deformation indeed occur, and is it reasonable to assume
gp$_{\delta \beta} \sim {\rm gp}_e$? (iii) How does the charging energy
affect our mechanism?

(i) To construct a more realistic case without the regularity of
Figs.~1 and 2, we consider the two-dimensional single-particle
Hamiltonian   
\begin{equation} 
\label{spham} 
H={{\bf p}^2\over 2m}+{m\omega^2\over 2} (x^2+(1-\beta
)^2y^2)-\lambda \hbar \omega L^2 ,
\end{equation}
where $L=i(y{\partial \over \partial x}-x{\partial \over \partial y})$
is the dimensionless $z$-component of the angular momentum operator. 
In nuclear physics, the three-dimensional analogue of $H$ is known as
the Nilsson model and has been quite successful in explaining the
spectra of deformed nuclei \cite{bm}. For $\beta > 0$ and $\lambda
\neq 0$, $H$ is not integrable \cite{hena4} and displays level
repulsion. We may view $H$ as a mean-field approximation to the total
Hamiltonian. The latter contains both, the mutual interaction of the
electrons and the real confining potential. Any such approximation
should lead to more or less chaotic single-particle motion~\cite{hena5}.

\begin{figure}
\epsfxsize=2.8in
\centerline{
\epsffile{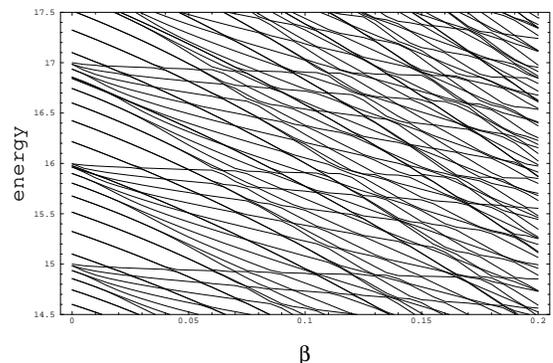}
%\epsffile{wei3.eps}
}
\vglue 0.1cm
\caption{
Energy levels as a function of $\beta $ of the Hamiltonian
 (\protect\ref{spham}) for $\lambda =0.004$.
}
\label{fig3}
\end{figure}

Fig.~3 shows part of the spectrum of $H$ versus $\beta $. For $\beta
=0$ the eigenvalues $m$ of $L$ are good quantum numbers. The levels
with $m \ne 0$ are pairwise degenerate. They form ascending sequences
with $\pm m, \pm(m-2),\ldots,\pm 1$ or 0. For $\beta>0$, each
eigenfunction contains a superposition of $m$-values. Depending on the
strength $\lambda$, the mixing may extend over $m$-values from
different shells. The overall pattern is quite similar to Fig.~1. The
resolution in Fig.~3 is not fine enough to show that all crossings are
avoided crossings. But is it realistic to assume that wave functions
on (nearly) horizontal levels retain their identity when $\beta$ is
changed? This, after all, is the central point used in our
argument. We have inspected the eigenvectors for the case $\lambda 
=0.005$ in the basis defined by the eigenvectors of the undeformed
Hamiltonian. The levels with the smallest slope have a dominant
component. It is the state with quantum numbers $n_y=0$ and
$n_x=N_{{\rm shell}}$. Here, $N_{{\rm shell}}$ denotes the shell
number. The modulus of the amplitude of this component 
is 0.85 or larger and remains virtually unchanged for $\beta \le
0.2$. Admixtures of non-dominant components with amplitude moduli
greater than 0.1 are limited to a small number of states (ten or
less). In contradistinction, the eigenvectors of the steepest levels
are genuine mixtures without dominant components. For $\beta > 0.1$,
the number of essential components (amplitude moduli larger than 0.1)
may be close to one hundred. There is a gradual transition, of course,
from the levels with the smallest to those with the largest slope.

The wave functions of the dominant components correspond classically
to orbits which oscillate in the $x$-direction, with little or no
motion in the $y$-direction. The relative stability of these orbits
is reflected in the small amount of mixing of their quantum analogues. 
We expect the bulk of the current to be carried by such orbits. The
experimental set-up appears to enforce occupation of the associated
characteristic levels as the conducting agent; otherwise a current
could barely flow. 

\begin{figure}
\epsfxsize=3in
\centerline{
%\epsffile{wei4.eps}
\epsffile{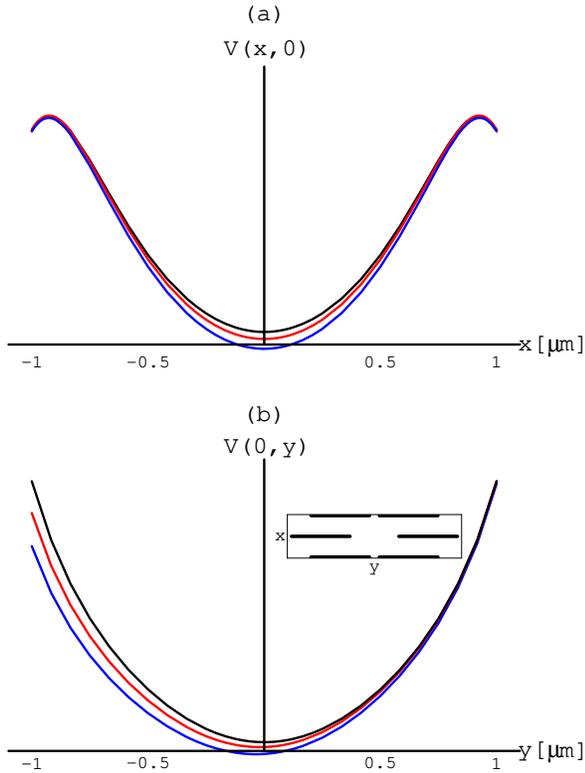}
}
\vglue 0.2cm
\caption{
Sections of the confining potential in (a) the $x$-direction and (b)
the $y$-direction in a plane parallel to the surface at a distance of
0.01 $\mu$m. The rods on the surface are arranged as shown in the
inset. The distance from top to bottom rod is 2 $\mu$m, the drawing is
in scale. The three curves in (a) and (b) refer to a plunger potential
 (middle left rod) of $-V_0,-0.9V_0$ and $-0.8V_0$, respectively, where
$-V_0$ is the potential at the other rods.
}
\label{fig4}
\end{figure}

(ii) Is it reasonable to discuss deformation, and how does variation
of the plunger gate voltage translate into a change of $\beta$? A
quantitative answer would require precise knowledge of a particular
experimental set-up, and we must confine ourselves to semiquantitative
considerations. We consider a set of gates arranged as indicated in
the inset of Fig.~4b. The current flows in the vertical direction. The
middle left conducting rod serves as plunger gate. To study the
potential well $V(x,y)$ formed beneath the gates, we have taken a set
of line charges. This avoids solving the full boundary-value problem. 
The charges on the top, middle right and bottom rods were all chosen
equal. Upon changing the charge on the plunger. they were re-adjusted
to ensure that the potential values at the height of the entrance and
exit barriers and at the tip of the middle right rod remain unchanged. 
An increase of the plunger voltage (charge) by 20\% results in a
decrease of 5\% and 0.7\% at the left hand and the right hand rods,
respectively. Sections of the potential well confining the electrons
taken in a plane below the gates are shown in Fig.~4, together with
the changes due to an increase of the plunger voltage. In the
$x$-direction, the bottom of the potential is lowered. In the
$y$-direction, the same lowering is accompanied by a widening toward
the left of the potential. Figure~4 indicates clearly the importance
of deformation.

To describe these changes quantitatively, we approximate the potential
by a paraboloid. The lowering of the bottom results in an effective
increase of $\omega_x$. In the $y$-direction, there is an effective
decrease of $\omega_y$. To first order in $\alpha$, we quantify these
changes as $\omega _x\to (1+\alpha)\omega_x$ and $\omega _y\to
(1-\alpha)\omega _y$. Rescaling the energy by a factor $(1+\alpha)$ and
writing $\beta = 2 \alpha$, we obtain $\omega_y \to (1-\beta)\omega
_y$ while $\omega _x$ is fixed. This is the situation discussed above. 
With $-V_0<0$ the original value of the voltage on all the gates, an
increase of the plunger potential from $-V_0$ to $-0.75V_0$ (a
realistic range of values) causes, in the example in Fig.~4, a
deformation $\beta \approx 0.1$. This is a sizeable value and
corresponds to the scales shown in Figs.~1 and 3.  

To relate gp$_{\delta \beta}$ and gp$_e$, we estimate the change of
depth of the harmonic oscillator potential needed to pull another
electron into the dot. As shown in Fig.~4a, we require that the values
of $V(\pm x_0, 0)$ remain unchanged under changes of the plunger
voltage. Here, $\pm x_0$ denote the $x$-values of the positions of the
barriers which define entrance and exit of the dot. We take $V(\pm
x_0, 0) = E_F$. (Since there are barriers, the value of $E_F$ is
actually slightly lower, but this is immaterial for what follows). 
In the case of Coulomb blockade without deformation \cite{been},
addition of one electron requires the potential to be lowered by
$\Delta_s +e^2/C \approx e^2/C$. Here, $\Delta_s$ is the average
single-particle level spacing and $e^2/C$ the charging energy. For
$N$ electrons, this yields $N e^2/C$ \cite{small}. Measured from the
points $\pm x_0$, the parabolic potential in $x$-direction must
therefore have the depth $(1+\alpha)^2 (\omega_x x_0)^2/2 \approx E_F +
Ne^2/C$.  Inclusion of deformation modifies the value of $C$ but does
not affect this argument qualitatively. We combine this result with the 
corresponding expression for the case of $N+1$ electrons at
deformation $\alpha + 1/2 \delta \beta$ to obtain $\delta \beta \sim
2 (e^2/C)/ (\omega_x x_0)^2 \sim 1/N$. In the last step we have
neglected $E_F$ in comparison with $Ne^2/C$. We compare this result
with the average distance in $\beta$ of nearest avoided crossings.
We consider the number $L$ of crossings on a horizontal level at $N_{{\rm
shell}}$ in Fig.~1. We find that $L$ is governed by $[\beta / (1 -
\beta)] N_{{\rm shell}}^2 /2$. For the distance of crossings this
yields $\delta \beta \sim 1 /N_{\rm shell}^2 \sim 1 / N$. This result
is generic. The estimates show that roughly one level crossing occurs
for each additional electron pulled into the  dot. This result lends
strong credibility to our model. 

(iii) Figs.~1 to 3 represent the single-particle energies without
taking account of the charging energy $U_0 = e^2/C$. To redress this
omission, we use the framework of the self-consistent treatment of the
Coulomb interaction within the quantum dot as applied in Ref.\ \cite{hawe}.
In the ordinary mechanism of avoided crossings, two levels
$\epsilon_j(V)$ with $j = 1,2$ are functions of an external parameter
$V$. Let the two levels be coupled by a (real) matrix element $W$. The
eigenvalues are 
\begin{equation}
\label{eig}
E_{\pm} = \frac{\epsilon_1 + \epsilon_2}{2}
\pm \sqrt{\left( \frac{\epsilon_1 - \epsilon_2}{2} \right)^2 + W^2}.
\end{equation}
If the two levels $\epsilon_j(V)$ intersect at some value $V =V_0$,
then the eigenvalues $E_{\pm}$ show an avoided crossing at $V = V_0$. 
We now include the charging energy by considering two ``effective''
single-particle levels ${\overline E_{\pm}}$ defined as solutions of
the mean-field equations \cite{hawe} ${\overline E_{\pm}} = E_{\pm} +
U_0 \langle n_{\mp} \rangle $. Here the $\langle n_{\pm} \rangle$ are
determined self-consistently,
\begin{eqnarray}
\langle n_{\pm} \rangle = \frac{1}{\pi} \left[ {\rm arctan} \left(
\frac{2 ( E_F - {\overline E_{\pm}})}{\Gamma_{\pm}} \right) + \frac{\pi}{2} 
\right] ,
\end{eqnarray}
where $\Gamma_{\pm}$ are the intrinsic widths of the two levels. We
solve these equations self-consistently for ${\overline E_{\pm}}$
and $\langle n_{\pm} \rangle$ and pick the solution that minimizes 
the mean-field energy  $E_{MF} = \langle n_{+} \rangle {\overline
E_{+}}+ \langle n_{-} \rangle  {\overline E_{-}} - q \langle n_{+}
\rangle \langle n_{-}  \rangle$. We focus on a situation where the
two effective levels are within an interval $U_0$ around the Fermi
energy, with one level below $E_F$ and the other above it. As a result
of the charging energy, the two levels participating in the avoided
crossing are simply shifted apart by roughly $U_0$. The wave function
at $V \ll V_0$ of the lower (higher) state is the same as the wave
function at $V \gg V_0$ of the higher (lower) state. The mixing of the
two wave functions near the avoided crossing is independent of the
charging energy $U_0$. As a result, the wave function of the state
that is occupied for $V \ll V_0$ changes smoothly into the wave
function of the state which is empty, and available for the next
electron, for $V \gg V_0$. This is in keeping with the mechanism
proposed above.

Clearly, strong correlations must be expected for the conductance peak
heights and for the transmission phases within a sequence of
resonances caused by one and the same quantum state. Long sequences of
this type are more likely to occur in quantum dots with regular shape. 
This might explain why long sequences of similar resonances where
observed for the symmetric (regular) dot of Ref.\ \cite{schu} and
perhaps weaker correlations in the more irregular dot of Ref.\
\cite{folk}. Our results suggest that further correlations not tested
so far in Ref.\ \cite{folk} might exist. This possibility as well
as a study of the effects of an applied magnetic field will be investigated
in a forthcoming paper. 

GH\ was supported by the Alexander von Humboldt Foundation and by
NSF Grant No.\ DMR-9215065. HAW was recipient of a South African/
Alexander von Humboldt research award when most of this work was
done. He acknowledges the warm hospitality of members of the
Department of Physics at Wits.

\end{multicols}


\begin{references}
\bibitem{small} The argument does not apply for $N$ of order 10 
or less where a constant-capacitance model fails. Experiments
typically use $N \approx 100-200$.
\end{references}
\end{document}